\PassOptionsToPackage{unicode}{hyperref}
\PassOptionsToPackage{hyphens}{url}
\PassOptionsToPackage{dvipsnames,svgnames,x11names}{xcolor}
\documentclass[
]{article}
\usepackage{amsmath,amssymb}
\usepackage{lmodern}
\usepackage[T1]{fontenc}
\usepackage[utf8]{inputenc}
\usepackage{textcomp} 
\IfFileExists{upquote.sty}{\usepackage{upquote}}{}
\IfFileExists{microtype.sty}{
  \usepackage[]{microtype}
  \UseMicrotypeSet[protrusion]{basicmath} 
}{}
\makeatletter
\@ifundefined{KOMAClassName}{
  \IfFileExists{parskip.sty}{%
    \usepackage{parskip}
  }{
    \setlength{\parindent}{0pt}
    \setlength{\parskip}{6pt plus 2pt minus 1pt}}
}{
  \KOMAoptions{parskip=half}}
\makeatother
\usepackage{xcolor}
\NewDocumentCommand\citeproctext{}{}
\NewDocumentCommand\citeproc{mm}{%
  \begingroup\def\citeproctext{#2}\cite{#1}\endgroup}
\makeatletter
 \let\@cite@ofmt\@firstofone
 \def\@biblabel#1{}
 \def\@cite#1#2{{#1\if@tempswa , #2\fi}}
\makeatother
\newlength{\cslhangindent}
\newlength{\csllabelwidth}
\newenvironment{CSLReferences}[2] 
 {\begin{list}{}{%
  \setlength{\itemindent}{0pt}
  \setlength{\leftmargin}{0pt}
  \setlength{\parsep}{0pt}
  \ifodd #1
   \setlength{\leftmargin}{\cslhangindent}
   \setlength{\itemindent}{-1\cslhangindent}
  \fi
  \setlength{\itemsep}{#2\baselineskip}}}
 {\end{list}}
\usepackage{calc}

\usepackage[bidi=default]{babel}
\babelprovide[main,import]{american}

\def\languageshorthands#1{}
\IfFileExists{bookmark.sty}{\usepackage{bookmark}}{\usepackage{hyperref}}
\IfFileExists{xurl.sty}{\usepackage{xurl}}{} 
\hypersetup{
  pdftitle={SWIFTGalaxy: a Python package to work with particle groups
from SWIFT simulations},
  pdfauthor={Kyle A. Oman},
  pdflang={en-US},
  colorlinks=true,
  linkcolor={Maroon},
  filecolor={Maroon},
  citecolor={Blue},
  urlcolor={Blue},
  pdfcreator={LaTeX via pandoc}}

\title{SWIFTGalaxy: a Python package to work with particle groups from
SWIFT simulations}

\definecolor{c53baa1}{RGB}{83,186,161}
\definecolor{c202826}{RGB}{32,40,38}

\usepackage[affil-it]{authblk}
\usepackage{orcidlink}
\author[1,2%
  \ensuremath\mathparagraph]{Kyle A. Oman%
    \,\orcidlink{0000-0001-9857-7788}\,%
    }

\affil[1]{Institute for Computational Cosmology, Physics Department,
Durham University, United Kingdom%
  }
\affil[2]{Centre for Extragalactic Astronomy, Physics Department, Durham
University, United Kingdom%
  }
\affil[$\mathparagraph$]{Corresponding author: %
}
\date{4 August 2024}

\begin{document}
\maketitle

\section{Summary}\label{summary}

SWIFTGalaxy is an open-source astrophysics module that extends
SWIFTSimIO (\citeproc{ref-Borrow2020}{Borrow \& Borrisov, 2020}) to
analyses of particles belonging to individual galaxies simulated with
SWIFT (\citeproc{ref-Schaller2024}{Schaller et al., 2024}). It inherits
from and extends the functionality of SWIFTSimIO's SWIFTDataset class.
It understands the content of halo catalogues and therefore which
particles belong to a galaxy or other group of particles, and its
integrated properties. The particles occupy a coordinate frame that is
enforced to be consistent, such that particles loaded on-the-fly will
match e.g.~rotations and translations of particles already in memory.
Intuitive masking of particle datasets is also enabled. Utilities to
make working in cylindrical and spherical coordinate systems more
convenient are also provided. Finally, tools to iterate efficiently over
multiple galaxies are provided.

\section{Background}\label{background}

Cosmological hydrodynamical galaxy formation simulations begin with a
representation of the early universe where its two key
constituents---dark matter and gas---are discretized. The dark matter is
usually represented with collisionless particles that respond only to
the gravitational force, while gas may be represented by particles
(Lagrangian tracers) or mesh cells (Eulerian tracers) and obeys the laws
of hydrodynamics in addition to gravity. The SWIFT code
(\citeproc{ref-Schaller2024}{Schaller et al., 2024}) takes the former
approach, implementing the smoothed particle hydrodynamics (SPH,
\citeproc{ref-Gingold1977}{Gingold \& Monaghan, 1977};
\citeproc{ref-Lucy1977}{Lucy, 1977}) formalism. Observations of the
cosmic microwave background are used to constrain a multiscale Gaussian
random field to formulate the initial conditions for a simulation of a
representative region of the early Universe, before the first galaxies
formed (e.g. \citeproc{ref-Bertschinger2001}{Bertschinger, 2001}). In
addition to hydrodynamics and gravity, galaxy formation models include
additional physical processes such as radiative gas cooling, star
formation, energetic feedback from supernovae and supermassive black
hole accretion, and more. Many of these are formally unresolved, leading
these to be collectively referred to as ``sub-grid'' elements of the
models.

As time integration of a cosmological hydrodynamical galaxy formation
simulation proceeds, outputs containing a ``snapshot'' of the simulation
state are typically written at regular intervals. These contain tables
of the properties of particles including at least positions, velocities,
masses and unique identifiers. More physically complex particle types,
such as stars and gas, carry additional properties such as temperature,
metallicity, internal energy, and many more. SWIFT snapshot files
include rich metadata recording the full configuration of the code and
galaxy formation model, plus physical units and related information for
all data.

For many types of analysis the snapshot files alone are insufficient:
knowledge of which groups of particles ``belong'' to gravitationally
bound structures---galaxies---is needed. In SWIFT this is most commonly
determined in a two-stage process. In the first stage particles are
linked to neighbouring particles separated by less than a suitably
chosen linking length to form connected ``friends-of-friends'' (FOF,
\citeproc{ref-Davis1985}{Davis et al., 1985}) groups. The second step
consists of finding overdense, self-bound particle groups or ``halos''
within each FOF group. There are numerous algorithms that accomplish
this task. The current implementation recommended for the SWIFT
community is the HBT-Herons (\citeproc{ref-ForouharMoreno2025}{Forouhar
Moreno et al., 2025}) implementation of the hierarchical bound-tracing
plus (HBT+, \citeproc{ref-Han2018}{Han et al., 2018}) algorithm. The
properties of halos can be tabulated in ``halo catalogues'', where again
many algorithms exist. This is often done as part of the halo-finding
process, but the recommendation for the SWIFT community is to move this
to a separate step and use the spherical overdensity and aperture
processor (SOAP, \citeproc{ref-McGibbon2025}{McGibbon et al., 2025})
tool.

\section{Statement of need}\label{statement-of-need}

The collection of standard cosmological hydrodynamical simulation data
products---snapshots of particles, halo membership information and halo
catalogues---leads to a generic workflow to begin analysis of individual
galaxies in such simulations. First, an object of interest is identified
in the halo catalogue. Then, the halo membership information is queried
to locate its member particles. Finally, the particles are loaded from
the snapshot to proceed with analysis. The SWIFT community already
maintains the SWIFTSimIO tool (\citeproc{ref-Borrow2020}{Borrow \&
Borrisov, 2020}) that supports reading in SWIFT snapshot files and SOAP
catalogue files. It uses metadata in the files to annotate data arrays
with physical units and relevant cosmological information. It is also
able to efficiently select spatial sub-regions of a simulation by taking
advantage of the fact the particles in cells of a ``top-level cell''
grid covering the simulation domain are stored contiguously. Finally, it
includes some data visualisation tools.

SWIFTGalaxy extends SWIFTSimIO by implementing the workflow outlined
above supplemented with additional features including coordinate
transformations, data array masking and efficient iteration over
galaxies. All of the SWIFTSimIO features, including use of the
visualisation tools, are inherited by SWIFTGalaxy. The package prevents
duplication of effort by users who would otherwise each need to
implement the same steps independently. It also helps to avoid common
errors, such as applying a coordinate transformation to one particle
type (e.g.~rotation of the dark matter) but forgetting to apply it to
others (e.g.~the gas remains unrotated) by enforcing a consistent data
state where applicable.

A core design principle of SWIFTGalaxy is that only operations with an
unambiguous implementation are included. In other words, SWIFTGalaxy
tries to avoid making decisions for its users. This is a key difference
when compared to other packages serving this purpose, such as pynbody
(\citeproc{ref-Pontzen2013}{Pontzen et al., 2013}) and yt
(\citeproc{ref-Turk2011}{Turk et al., 2011}) that can also be used with
SWIFT data sets. Other packages are also less tailored to SWIFT and
therefore less able to take advantage of the detailed structure of the
data layout on disk and information available as metadata.

Despite being tailored for use with SWIFT simulations, SWIFTGalaxy is
modular: it has interchangeable interfaces to different halo catalogue
formats. In addition to the preferred SOAP format, the Velociraptor
(\citeproc{ref-Elahi2019}{Elahi et al., 2019}) and
\href{https://github.com/dnarayanan/caesar}{Caesar} formats are also
supported, and adding support for other formats is straightforward by
design. A ``standalone'' mode is also available so that the other
features of SWIFTGalaxy can be used on arbitrary collections of
particles even if no halo catalogue is available or desired.

SWIFTGalaxy is hosted on
\href{https://github.com/SWIFTSIM/swiftgalaxy}{GitHub} and has
documentation available through
\href{https://swiftgalaxy.readthedocs.io}{ReadTheDocs}. The first
research article using the package has recently appeared
(\citeproc{ref-Trayford2025}{Trayford et al., 2025}). Many more projects
using it are currently ongoing, especially in the context of the COLIBRE
simulations project (\citeproc{ref-Chaikin2025}{Chaikin et al., 2025};
\citeproc{ref-Schaye2025}{Schaye et al., 2025}).

\section{Acknowledgements}\label{acknowledgements}

KAO acknowledges support by the Royal Society through Dorothy Hodgkin
Fellowship DHF/R1/231105, by STFC through grant ST/T000244/1, and by the
European Research Council (ERC) through an Advanced Investigator Grant
to C. S. Frenk, DMIDAS (GA 786910). This work has made use of NASA's
Astrophysics Data System.

\section*{References}\label{references}
\addcontentsline{toc}{section}{References}

\phantomsection\label{refs}

\end{document}